\documentclass[twocolumn,amstext,amsmath,amssymb,prl,superscriptaddress,floatfix,showpacs,nofootinbib]{revtex4}

\usepackage[dvips]{epsfig}
\usepackage{slashed}
\usepackage{mathrsfs}
\usepackage{graphicx}
\usepackage{xcolor}
\usepackage{hyperref}
\usepackage{dcolumn}
\usepackage{subfigure}
\usepackage{xspace}

\usepackage{color}
\usepackage{ulem}

\newcommand{\Eqref}[1]{Eq.~\eqref{#1}} 
 
\newcommand{\mm}{\marginpar{\colorbox{green}{\textbf{BJ}}\\@Mario:}}
\newcommand{\bj}{\marginpar{\colorbox{green}{\textbf{MM}}\\@BJ:}}

\def\roughly#1{\mathrel{\raise.3ex\hbox{$#1$\kern-.75em%
\lower1ex\hbox{$\sim$}}}}

\def\g2k{\Gamma^{(2)}_k}

\def\ma0{m_{a_{0}}}
\def\mf0{m_{f_{0}}}

\definecolor{heidelbeer}{rgb}{0.5,0,0.5}

\graphicspath{
{./}
{../figures/}
}
\begin{document}
\title{Effective Mass Signatures in Multiphoton Pair Production}

\author{Christian Kohlf\"{u}rst}
 \email[]{christian.kohlfuerst@uni-graz.at}
 \affiliation{Institut f\"{u}r Physik, Karl-Franzens-Universit\"{a}t,
   A-8010 Graz, Austria}
  \affiliation{Theoretisch-Physikalisches Institut, Abbe Center of Photonics, Friedrich-Schiller-Universit\"{a}t Jena, D-07743 Jena, Germany}

  \author{Holger Gies}
 \email[]{holger.gies@uni-jena.de}
  \affiliation{Theoretisch-Physikalisches Institut, Abbe Center of Photonics, Friedrich-Schiller-Universit\"{a}t Jena, D-07743 Jena, Germany}
\affiliation{Helmholtz-Institut Jena, Fr\"obelstieg 3, D-07743 Jena, Germany}

  \author{Reinhard Alkofer}
 \email[]{reinhard.alkofer@uni-graz.at}
  \affiliation{Institut f\"{u}r Physik, Karl-Franzens-Universit\"{a}t,
   A-8010 Graz, Austria}

\date{\today}

\begin{abstract}
{Electron-positron pair production in oscillating electric fields
is investigated in the nonperturbative threshold regime. 
Accurate numerical solutions of quantum kinetic theory for 
corresponding observables are presented and analyzed in terms
of a proposed model for an effective mass of electrons and positrons acquired
within the given strong electric field. Although this effective mass 
cannot provide an exact description of the collective
interaction of a charged particle with the strong field,
physical observables are identified which carry direct and
sensitive signatures of the effective mass.}
\end{abstract}

\maketitle

\textit{Introduction.}--The concept of a particle's effective mass $m_\ast$ is widely used
in physics. It typically describes the mass a particle seems to have
when responding to an external probe while it is immersed in an
ambient medium or thermal bath. The value $m_\ast$ therefore
parameterizes the consequences of a particle's collective interactions
with its environment in a simple way. The effective mass concept hence
occurs not only in many branches of condensed matter physics, but also,
{\it e.g.,} 
in the form of constituent masses of quarks inside a hadron. Even
the fundamental fermion masses of the standard model may be viewed as
effective masses induced by the interactions with the Higgs field.

While obviously useful, the concept of effective masses may be
questioned, since the reduction of collective interactions into
a single number $m_\ast$ can be oversimplifying, and the actual value
of $m_\ast$ can depend on many details of an actual experimental
setup and thus be generically nonuniversal. A comparatively clean
setting in which effective masses have been discussed is given by the
dynamics of electrons in a strong electromagnetic field. In
particular, in periodically modulated fields the analogy to
condensed-matter systems appears immediate.

Phenomena which are describable by an effective mass acquired by the
electron in a strong field have been studied in undulator fields used
for free electron lasers \cite{mStar:FELs} as well as in plane wave
fields \cite{mStar:Lasers}. Also, concepts for general fields
  have been suggested \cite{Dodin:2008}. In particular,
  generalizations from the idealized plane-wave case to realistic
  laser fields have lead to a diversified discussion in the literature
  \cite{Kibble:1966zza,Neville:1971,Boca:2009zz,Heinzl:2009nd,Seipt:2010ya,Mackenroth:2010jr,Corson+Reply:2012}.
Recently a concrete setup has been proposed \cite{Harvey:2012ie}, how
effective-mass effects could be observed in laser-electron scattering
experiments with current technology. It was also demonstrated
explicitly that the effective mass is not universal but depends, for
instance, on laser pulse shapes.

All these studies have in common that the effective mass is not
observed by looking directly at the produced electrons and positrons, but can be deduced from properties of the photon
emission spectra arising from the electron dynamics. As the particle
spectrum does not show a heavy electron excitation \cite{Ilderton:2012qe}, in these processes
the effective mass occurs as an auxiliary quantity at intermediate
stages of the calculation. More precisely, it arises as the invariant
square of an averaged (quasi-)four-momentum associated with the
electron motion inside the periodic field. The effective mass in a monochromatic plane wave, for example,
is (in lightcone gauge) given by \cite{Wolkow:1935zz}
\begin{align}
  m_* = m \sqrt{1 + \xi^2} \ , \ \textrm{where} \ \xi = \frac{e}{m} \sqrt{- \langle A^{\mu} A_{\mu} \rangle } \ .
  \label{eq1}
\end{align}

The generalization to arbitrary pulses is given in
  \cite{Kibble:1975vz}. The present work is devoted to investigate
whether the effective mass concept in strong fields might be more
directly accessible in terms of a process which shows a severe
sensitivity to a mass threshold {\cite{Burke:1997ew,Bamber:1999zt}}.
For this, we propose the characteristic phenomenon of pair production
in strong electric fields where electromagnetic energy is converted
into the creation of electrons and positrons
\cite{Sauter:1931zz,Heisenberg:1935qt,Schwinger:1951nm}. To be
  more specific, we consider only pair production directly from
  colliding laser pulses in a regime with both multiphoton and
  nonperturbative features. Since the phase space available for pair
creation is extremely sensitive to the masses of the particles to be
generated, quantities such as the particle yield should give a rather
direct access to the effective mass. The fact that the effective mass
  of \Eqref{eq1} governs the threshold for pair production in laser
  fields has already been noted in
  \cite{Harvey:2012ie,Heinzl:2010vg} in the context of
  stimulated pair production \cite{Schutzhold:2008pz} and in
  \cite{PhysRevA.81.022122} for oscillating electric fields, see
    also \cite{Titov:2012rd,Nousch:2012xe} for short pulses.

In order to quantify the influence of the effective mass
  concept, we study (multiphoton) pair production in rapidly
oscillating electric fields close to the pair production threshold by
means of quantum kinetic theory (QKT), requiring highly accurate
  numerical solutions. As an advantage of QKT, quantum
  statistical effects as well as non-Markovian memory effects are
  included
  \cite{Smolyansky:1997fc,Kluger:1998bm,Schmidt:1998vi}. Also, the
  real-time evolution of distribution functions is directly accessible
  in this framework. This gives us access not only to the total
particle yield, but also to the momentum distributions of the outgoing
pairs. From the latter, we are able to read off the
characteristics of above-threshold phenomena similar to above-threshold ionization (ATI) 
spectra in the context of atomic ionization
{\cite{Delone:1994}. Similarities between atomic ionization and
  pair production have been known for a long time
  \cite{Popov:1973}. They become conceptually obvious in path integral
  approaches \cite{Salieres:2001,Dunne:2005sx}, and have first been
  quantified for the case of superimposed laser and Coulomb fields as
  well as counterpropagating lasers in \cite{Muller:2009zzf}.

In this Letter, we consider comparatively simple field
  configurations such as oscillating electric fields. Still, we
  observe that the observables can develop a rather involved
  dependence on the underlying parameters. In particular, sufficiently
  accurate numerical solutions are necessary, to resolve the
  complexity of the pair-production process. Despite this complexity,
  we find that features of many observables can, in fact, be directly
understood in terms of an intuitive effective-mass description,
phrased in terms of a simple model for $m_\ast$.

\textit{Electric field pulse.}--For our pair production studies, we 
use a homogeneous electric field pulse with peak field strength
$\varepsilon$, frequency $\omega$, and duration $\tau$ of the form
\begin{equation}
  E(t) = \varepsilon \ \mathrm{exp} \left( - \frac{t^2}{2\tau^2} \right) \ \mathrm{cos} \left(\omega t \right) \ .
  \label{eq3}
\end{equation}
Our region of interest, where some of these parameters are close to
the Compton scale, is currently not yet accessible by experiments, but
may come within reach with future tailored x-ray laser
beams {\cite{Ringwald:2001ib}. Equation \eqref{eq3} may be thought of as a model of the
  electric field in an antinode of a standing-wave mode ignoring
  spatial inhomogeneities. A more realistic modeling would require 
  including also possible magnetic components, along the lines of
  \cite{Ruf:2009zz,Muller:2009zzf} as well as variations in space and time
  \cite{Ruf:2009zz,Hebenstreit:2011wk}.}  The corresponding vector
potential reads
{
\begin{equation}
A(t) = -\frac{1}{2} e^{-\frac{1}{2} \omega^2 \tau ^2} \sqrt{\frac{\pi }{2}} \ 
\varepsilon \ \tau 
\left(\text{Erf}\left[\frac{t-i \omega \tau ^2}{\sqrt{2} \tau }\right]+c.c.\right).
\end{equation}
}
 
\textit{Quantum Kinetic Theory.}--Our following results for the
  pair production process are based on quantum kinetic theory QKT
  \cite{Smolyansky:1997fc,Kluger:1998bm,Schmidt:1998vi}. We emphasize
  that numerical accuracy is essential for some of our observations
  presented in the following. A central quantity in QKT is the one-particle
distribution function $F(q,t)$ which, when evaluated at asymptotic
times, allows to extract momentum spectra for $e^-$ and
$e^+$. Accordingly, the particle yield per unit volume and $d^2 q_{\perp}/(2\pi)^2$ is given by $N = \int
dq / (2 \pi) \ F(q,\infty)$. In the Weyl gauge, the Vlasov equations of QKT can
be written as \cite{Bloch:1999eu,Alkofer:2001ik}
 \begin{align}
 \dot{F}(q,t) & =  W(q,t) \ G(q,t) \ ,  \\
 \dot{G}(q,t) & =  W(q,t) \ [1-F(q,t)]-2 w(q,t) \ H(q,t) \ ,  \\
 \dot{H}(q,t) & =  2 w(q,t) \ G(q,t) \ , 
 \end{align}
where
\begin{equation}
 w^2(q,t) = m^2 + [q - e A(t)]^2 \ , \ W(q,t) = \frac{eE(t) m}{w^2(q,t)} \ .
\end{equation}
Here, $G(q,t)$ and $H(q,t)$ are auxiliary functions.  Throughout the
article, $q\equiv q_\|$ denotes the kinetic momentum parallel to the
electric field as we perform all calculations with $q_{\perp}=0$.  The
initial conditions are given by
$F(q,-\infty)=G(q,-\infty)=H(q,-\infty)=0$. A typical result for the
particle yield $N$ per Compton wavelength $\lambda_C$ below and
  near the pair threshold $\omega=2m$ for the example of
  $e\varepsilon/m^2=0.1$ and a comparatively long pulse of
  $\tau=100/m$ is shown in Fig.~\ref{fig:Distr}.

  \begin{figure}[t] 
  \centering
  \includegraphics[width=\columnwidth]{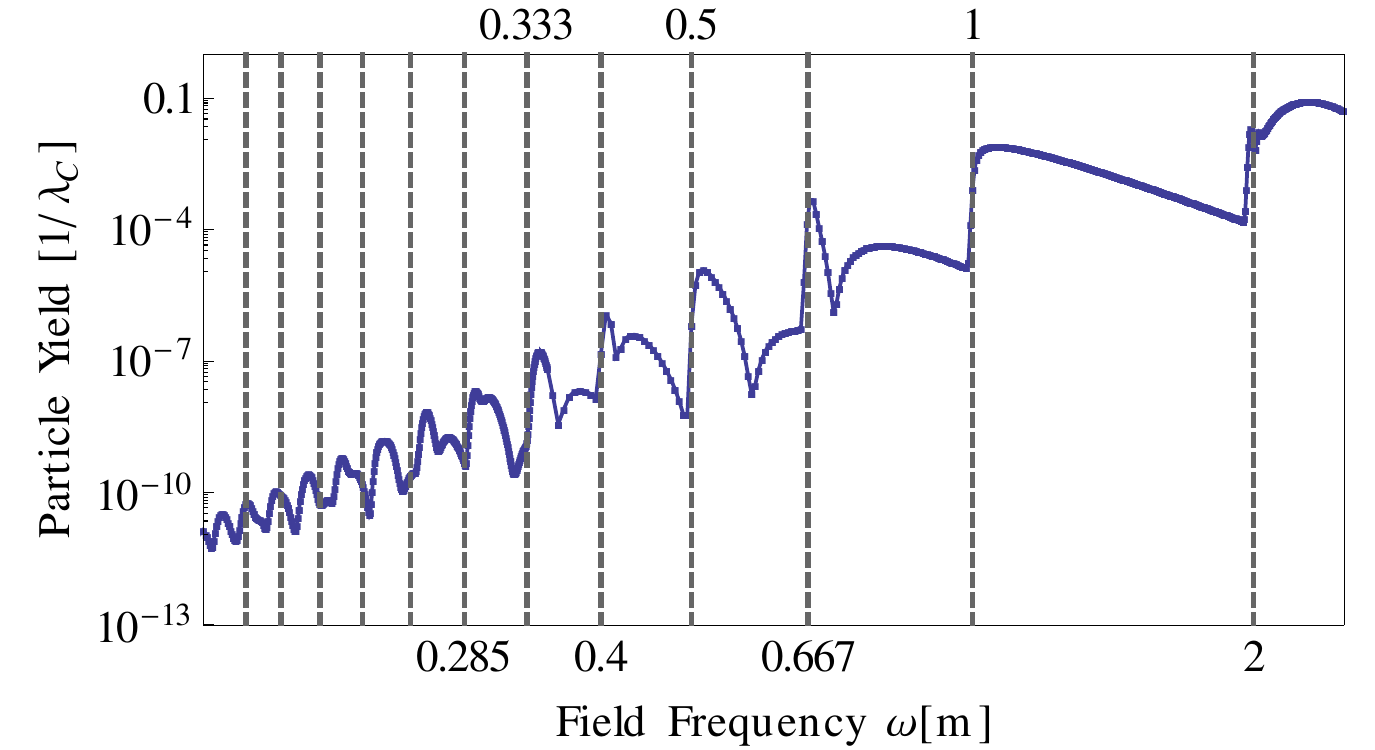}
  \caption{\label{fig:Distr} Double-log plot of the particle yield as
    a function of frequency $\omega$ for $\tau=100/m$ and
    $e\varepsilon/m^2=0.1$. The oscillating structure can be related to the
    $n$-photon thresholds. The true peak positions are typically
    slightly above the naive threshold estimate $n \omega =2 m$
    (dashed lines).}
    
\end{figure}
  
The particle yield exhibits a characteristic oscillatory structure
which can be interpreted as a signature for multiphoton
production. Similar observations based on the straightforward
  numerical solution of the Dirac equation for slightly different
  parameters have been made in \cite{PhysRevA.81.022122}. Naively,
the phase space for $n$-photon pair production is expected to open at
the threshold frequencies, satisfying $n \omega= 2 m$ (dashed lines in
Fig.~\ref{fig:Distr}). Upon close inspection, we find that the
particle yield is actually peaked slightly above this naive
expectation. This deviation  becomes more
pronounced for higher $n$. In the following we will present arguments
 that this deviation can be
interpreted as a signature for the effective mass of the
electron or positron in the strong field.

\textit{Effective mass model.}--  
Although \Eqref{eq1} holds strictly
only for plane-wave fields (possible generalizations have been
discussed in \cite{Ruf:2009zz,Harvey:2012ie}), we suggest to use
\Eqref{eq1} together with an average over one field oscillation inside
the pulse envelope. Confining ourselves to long pulse trains with
$\tau \gtrsim 100/m$ such that we can ignore the finite band width
of the pulse, we propose
\begin{equation}
  m_\ast = m \sqrt{1 + \xi^2} \approx m \sqrt{1 + \frac{e^2}{m^2} \frac{\varepsilon^2}{2 \omega^2} } \ .
\label{eq:effmass}
\end{equation}
This model suggests that the naive threshold frequencies for
$n$-photon production should be replaced by $n\omega = 2 m_\ast$.
This threshold condition together with \Eqref{eq:effmass} can be
resolved in terms of the threshold frequencies,
\begin{equation}
  \omega_n = \sqrt{ \frac{2m^2}{n^2} + \sqrt{ \frac{4 m^4}{n^4} + \frac{2 {e^2} \varepsilon^2 m^2}{n^2}}} \ ,
  \label{eq2}
\end{equation}
or, correspondingly, in terms of the  effective mass at the $n$th threshold
\begin{equation}
  m_{\ast,n} = \sqrt{ \frac{m^2}{2} + \frac{\sqrt{ 2m^4 + n^2 {e^2} \varepsilon^2 m^2} }{2 \sqrt{2}}} \ .
\label{eq:mastn}
\end{equation}
Equation \eqref{eq2} predicts that the threshold frequencies move to
larger values for increasing field strength. This is illustrated in
Fig.~\ref{fig:Yield_g7} for the (normalized) particle yield near the
heptaphoton $n=7$ threshold for $\tau =100/m$. For increasing field
strength $e\varepsilon/m^2=0.02, \dots, 0.2$ the shift of the peak towards
larger frequencies is clearly visible.

\begin{figure}[t]
  \centering
  \includegraphics[width=\columnwidth]{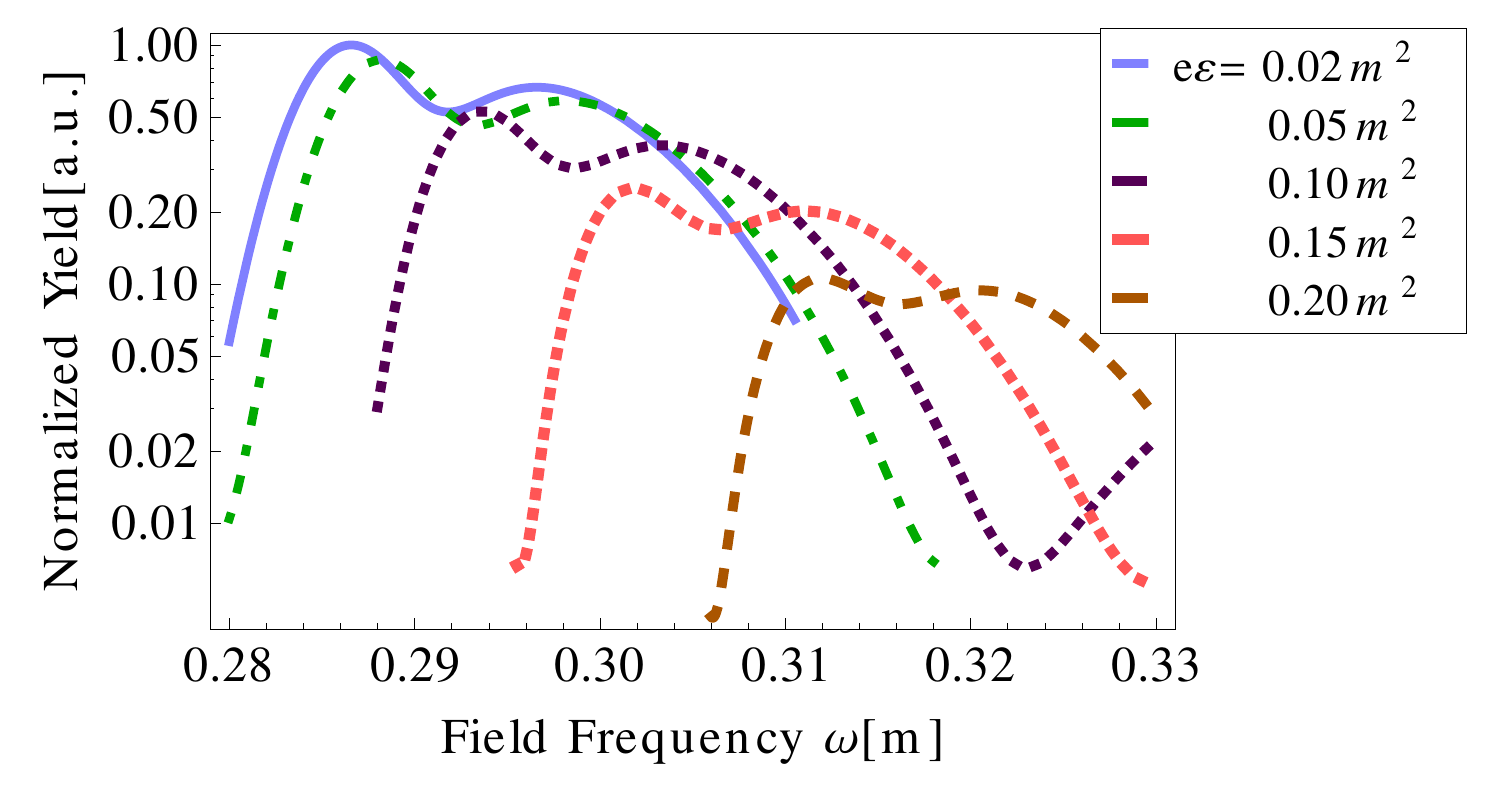}
  \caption{\label{fig:Yield_g7} Normalized ($n=7$) heptaphoton-production
    particle yield in a log-linear plot as a function of
    frequency for several values of the peak field strength
    $e\varepsilon/m^2=0.02, \ldots, 0.2$(from left to right). 
    The threshold frequency, being close to $\omega/m\simeq
    2/7 \simeq0.286$ for weak fields, moves to larger values for
    increasing field strength.  (The yield is normalized by a factor
    $\sim\varepsilon^{14}$ to account for the trivial multiphoton
    $\varepsilon^{2n}$ dependence).}
\end{figure}

A quantitative comparison between the full numerical solution of QKT
and the effective mass model \eqref{eq:mastn}
can be seen in Fig.~\ref{fig:EffMass}
again for the heptaphoton case ($n=7$) and $\tau=100/m$. Here, we
compare the effective mass $m_{\ast,7}=(7/2) \omega_{\text{peak}}$
obtained from the peak position $\omega_{\text{peak}}$ of the
heptaphoton yield (corresponding maxima of the curves in
Fig.~\ref{fig:Yield_g7}) with the model prediction \eqref{eq:mastn}. 
There is considerable agreement over a wide range of field
strengths showing an increase of the effective mass 
up to $15\%$ for the chosen parameters.

\begin{figure}[t]
  \centering
  \includegraphics[width=\columnwidth]{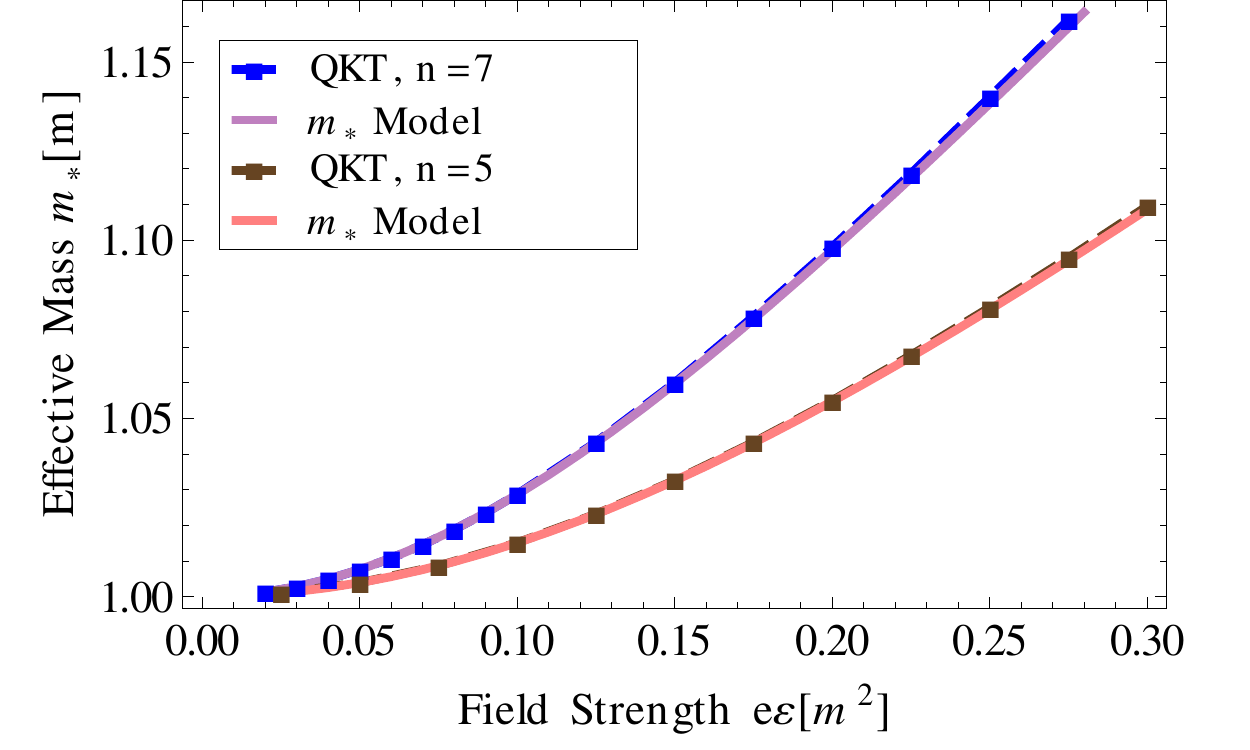}
  \caption{\label{fig:EffMass} Effective mass $m_{\ast,n=7}$(upper curves) and $m_{\ast,n=5}$(lower curves) extracted
    from the $n$-photon peak of the particle yield from full QKT
    solution in comparison with the effective mass model prediction
    \eqref{eq:mastn} as a function of the field strength $e\varepsilon/m^2$ 
    ($\tau=100/m)$. }
\end{figure}

\textit{Momentum distribution.}--Effective mass effects do not only
show up in the total particle yield. With the aid of QKT, we can also
determine the one-particle distribution in momentum space
$F(q,\infty)$. An example is given in Fig.~\ref{fig:FullYield} for
$\omega=0.3 m$, {$e\varepsilon=0.2m^2$} and $\tau=300/m$. In addition
to a broad peak with an involved substructure near $q=0$, we observe a
regular pattern of higher-momentum peaks. As these additional peaks
are separated by multiples of the frequency ($\omega=0.3m$ in the
present case), we interpret the peaks as signatures of the absorption
of additional photons in the pair production process. In other words,
an $n$-photon peak in the total particle yield
(cf. Fig.~\ref{fig:Distr}) receives contributions also from production
processes involving $n+s$ photons. The momentum distribution of
outgoing particles as in Fig.~\ref{fig:FullYield} serves as a
spectrometer for the production processes with $s=0,1,2, \dots$.  This
is similar to ATI spectra in the field of atomic ionization
\cite{Delone:1994}. This quantitative similarity to pair production
has also been noticed in \cite{PhysRevA.81.022122,Ruf:2009zz}.

\begin{figure}[h]
  \centering
  \includegraphics[width=\columnwidth]{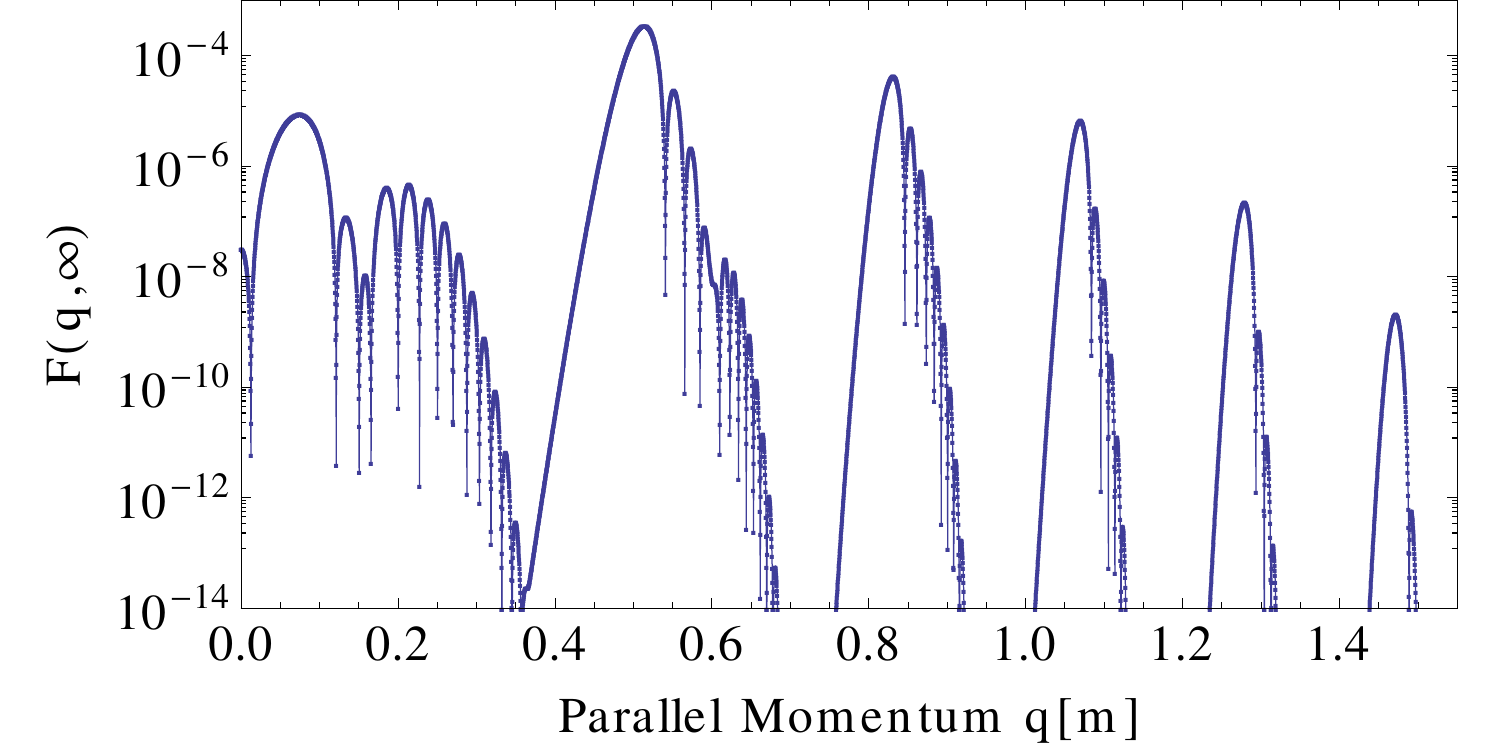}
  \caption{\label{fig:FullYield} Spectrum of the one-particle
    distribution for multiphoton pair production as a function of
    kinetic momentum.  The peaks for higher momenta arise due to the
    absorption of $s$ additional photons. Pulse parameter: $\tau=300
    /m$, $\omega=0.3 m$ and {$e\varepsilon=0.2m^2$}.}
\end{figure}

\textit{Channel closing.}--The similarity to ATI spectra gives rise to
another way for directly and sensitively observing effective mass
effects.  From energy conservation of multiphoton pair production,
$E_{(n+s)\gamma} = E_{e^-} + E_{e^+}$, we obtain the general relation
\begin{equation}
  \left( \frac{(n+s) \omega}{2} \right)^2 = m_{\ast}^2 + q_{n+s}^2 \ ,
\label{eq:chcl}
\end{equation}
where we have used the effective mass as a substitute for the naive
electron rest mass inside the electric field, and $q_{n+s}$ denotes a
characteristic residual kinetic momentum of the outgoing particles in
the momentum-space peak generated from the $n+s$ multiphoton
process. Let us fix the number of photons $n+s$ as well as the photon
frequency $\omega$. Since $m_{\ast}$ increases with the field
strength $\varepsilon$, cf. \Eqref{eq:effmass}, \Eqref{eq:chcl} predicts that the
characteristic momentum $q_{n+s}$ has to decrease with increasing
$\varepsilon$.

This effect can indeed be observed in our full QKT results: In
Fig.~\ref{fig:Closing}, we plot the peak position of the $(n+s=7)$
photon peak in the momentum distribution for a frequency
$\omega=0.322m $ as a function of the field strength $\varepsilon$ and
for a pulse duration $\tau=300/m$. For increasing field strength
$\varepsilon$, the peak position moves to lower momenta $q_{n+s}$ as
expected from \Eqref{eq:chcl}. 
(NB: The discontinuities in the peak position arise because the role of main and side maxima can interchange with increasing field strength,
cf.\ the substructure of the one-particle distribution displayed in 
Fig.~\ref{fig:FullYield} and the discussion below.)
Once, the effective mass exceeds the
available multiphoton energy, the corresponding momentum peak in the
distribution vanishes. This effect is known as ``channel closing'' in
atomic ionization \cite{Kobold:2002}.

More quantitatively, the approach to the threshold first leads to a
splitting of the momentum peak under consideration into several
  narrower peaks. To allow for a unique determination of the peak
  position we plotted in Fig.~\ref{fig:Closing} always the position of
  the highest local maximum which then in turn leads to the
  displayed discontinuities. At the threshold, the peaks
vanish. Pictorially speaking, the left-most peak in the corresponding
analog of Fig.~\ref{fig:FullYield} drops out of the plot window
towards the left edge. Again, the phenomenon of channel closing
observed in the full QKT data agrees rather well with the simple
effective mass model \eqref{eq:mastn}. Both, the overall shift of the
momentum peaks and the location of the threshold are nicely
described. We conclude that channel closing is an observable
  with a strong sensitivity to the effective mass.

\begin{figure}[t] 
  \centering
  \includegraphics[width=\columnwidth]{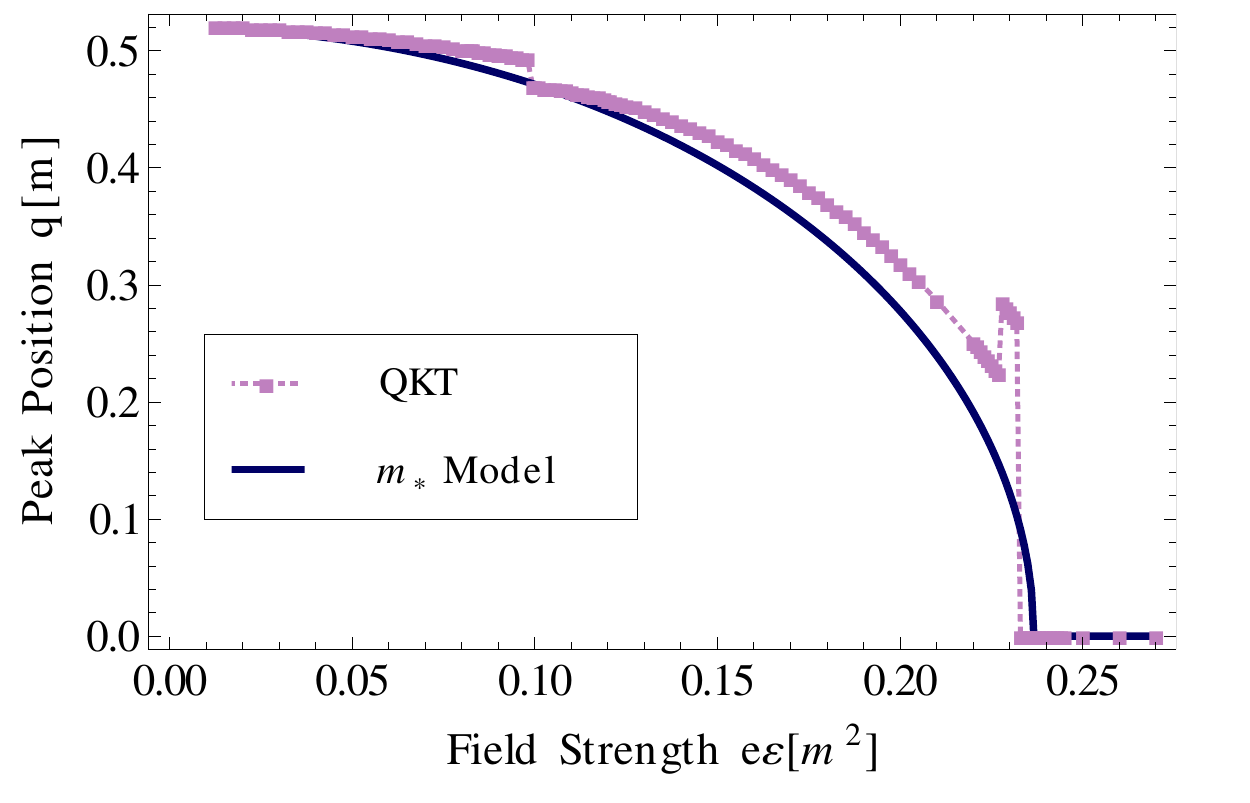}
  \caption{\label{fig:Closing} 
   Position of the $n+s=7$ 
photon peak in the momentum distribution as a function of field
strengths for a photon frequency of $\omega=0.322 m$ and pulse length
$\tau=300/m$. }
\end{figure}

\textit{Discussion.}--Given the complexity of pair production and
corresponding phase-space observables already for the case of such
simple field configurations as considered here, it is clear that the
simple effective mass model can only describe a limited set of
quantities. Nevertheless, it is worthwhile to stress that the
effective mass signatures are omnipresent in the investigated
parameter regime.

The limitations of the effective mass description become clear by
noting that the effective mass of \Eqref{eq:effmass} is related to the
Keldysh parameter {$\gamma_\omega=m\omega/(e \varepsilon)$} by $m_\ast=m
\sqrt{1+1/(2\gamma_\omega^2)}$. The Keldysh parameter distinguishes
the limit of nonperturbative Schwinger pair production
$\gamma_\omega\ll 1$ from the perturbative multiphoton regime for
$\gamma_\omega\gg 1$ \cite{Brezin:1970xf}. Our results have been
obtained in the regime $\gamma_\omega\gtrsim 1$ which characterizes
the nonperturbative threshold regime, exhibiting features of both
regimes -- nonperturbative pair production as well as multiphoton
processes. Towards the multiphoton regime $\gamma_\omega\gg 1$ the
effective mass approaches the vacuum mass rendering a distinction
irrelevant. For smaller $\gamma_\omega$ towards the Schwinger regime,
deviations from the simple effective mass model appear to
occur in our data. However, in this regime backreactions can become relevant
\cite{Bloch:1999eu,Hebenstreit:2013qxa} which are not included in our
description. 

The considered field pulses allow for a second
Keldysh parameter $\gamma_\tau = m/(\tau e \varepsilon)$ which is irrelevant for
the long pulses discussed here. For shorter pulses, the interplay
between $\tau$ and $\omega$ can lead to distinct momentum signatures
\cite{Hebenstreit:2009km}. 

Finally, we point out that the quantitative precision of the effective
mass model also differs in different parameter regimes. This can be
traced back to involved substructures of the momentum distributions:
for instance, for multiphoton production with even $n$, the momentum
distribution has to vanish at $q=0$ due to charge-conjugation
invariance \cite{PhysRevA.81.022122}. This takes a quantitative
influence on the total particle yield and hence also on the
quantitative agreement with the effective mass model. We emphasize
that
all qualitative effective mass dependencies are left untouched by
these details.

\textit{Summary.}--Based on accurate numerical solutions of quantum
kinetic theory for pair production in oscillating electric fields, we
have demonstrated that experimentally accessible observables show
clear traces of the effective mass of electrons and positrons acquired
within a strong electric field. Whereas the effective mass is neither
a universal quantity nor an exact description of the collective
interaction of a charged particle with the strong field, we have
identified physical observables that carry direct signatures of the
effective mass in a sensitive manner. From a pragmatic viewpoint, we
believe that this justifies to say that the effective mass ``can be
measured''.

\textit{Acknowledgements.}
We are grateful to A. Blinne for
interesting and enlightening discussions. C.K. is funded by the Austrian
Science Fund, FWF, through the Doctoral Program on Hadrons in Vacuum,
Nuclei, and Stars (FWF DK W1203-N16).  We thank the research core area
``Modeling and Simulation'' for support.  H.G. acknowledges support by
the DFG under Grants No. Gi 328/5-2 (the Heisenberg program) and No. SFB-TR18. \


\begin{thebibliography}{100} 

\bibitem{mStar:FELs}
Z.~Huang and K.-J.~Kim, Phys. Rev. ST Accel. Beams {\bf 10},
034801 (2007); B.~W.~J.~McNeil and N.~R.~Thompson, Nature Photonics {\bf 4}, 814 (2010).


\bibitem{mStar:Lasers}
D.~M.~Wolkow,
Z.~Phys.~ {\bf 94}, 250 (1935);
%
I.~I.~Goldman, Phys.\ Lett.\ {\bf 8}, 103 (1964); A.~I.~Nikishov and
V.~I.~Ritus, Sov.~Phys.~JETP {\bf 19}, 529 (1964); %
%
  T.~W.~B.~Kibble,
  Phys.\ Rev.\  {\bf 138}, B740 (1965).
%

\bibitem{Dodin:2008}
I.~Y.~Dodin and N.~J.~Fisch, Phys.\ Rev.\ E {\bf 77}, 036402 (2008).

\bibitem{Kibble:1966zza} 
  T.~W.~B.~Kibble,
  Phys.\ Rev.\  {\bf 150}, 1060 (1966).

\bibitem{Neville:1971}
R.~A.~Neville and F.~Rohrlich, Phys.\ Rev.\  D {\bf 3}, 1692 (1971).

\bibitem{Boca:2009zz} 
  M.~Boca and V.~Florescu,
  Phys.\ Rev.\ A {\bf 80}, 053403 (2009).


\bibitem{Heinzl:2009nd} 
  T.~Heinzl, D.~Seipt and B.~Kampfer,
  Phys.\ Rev.\ A {\bf 81}, 022125 (2010)

\bibitem{Seipt:2010ya} 
  D.~Seipt and B.~Kampfer,
  Phys.\ Rev.\ A {\bf 83}, 022101 (2011)


\bibitem{Mackenroth:2010jr} 
  F.~Mackenroth and A.~Di Piazza,
  Phys.\ Rev.\ A {\bf 83}, 032106 (2011)

\bibitem{Corson+Reply:2012}
J.~P.~Corson and J.~Peatross, Phys.\ Rev.\ A {\bf 85}, 046101 (2012);
F. Mackenroth and A. Di Piazza, 
Phys.~Rev.~A {\bf 85}, 046102 (2012).

\bibitem{Harvey:2012ie} 
  C.~Harvey, T.~Heinzl, A.~Ilderton and M.~Marklund,
  Phys.\ Rev.\ Lett.\  {\bf 109}, 100402 (2012)
{  
\bibitem{Ilderton:2012qe} 
  A.~Ilderton and G.~Torgrimsson,
  Phys.\ Rev.\ D {\bf 87}, 085040 (2013)
}
\bibitem{Wolkow:1935zz} 
  D.~M.~Wolkow,
  Z.\ Phys.\  {\bf 94}, 250 (1935).
{
\bibitem{Kibble:1975vz} 
  T.~W.~B.~Kibble, A.~Salam and J.~A.~Strathdee,
  Nucl.\ Phys.\ B {\bf 96}, 255 (1975).
}  
\bibitem{Burke:1997ew} 
  D.~L.~Burke, R.~C.~Field, G.~Horton-Smith, T.~Kotseroglou, J.~E.~Spencer, D.~Walz, S.~C.~Berridge and W.~M.~Bugg {\it et al.},
  Phys.\ Rev.\ Lett.\  {\bf 79}, 1626 (1997).
  
\bibitem{Bamber:1999zt} 
  C.~Bamber, S.~J.~Boege, T.~Koffas, T.~Kotseroglou, A.~C.~Melissinos, D.~D.~Meyerhofer, D.~A.~Reis and W.~Ragg {\it et al.},
  Phys.\ Rev.\ D {\bf 60}, 092004 (1999).
  
\bibitem{Sauter:1931zz} 
  F.~Sauter,
  Z.\ Phys.\  {\bf 69}, 742 (1931).
  
\bibitem{Heisenberg:1935qt} 
  W.~Heisenberg and H.~Euler,
  Z.\ Phys.\  {\bf 98}, 714 (1936)

\bibitem{Schwinger:1951nm} 
  J.~S.~Schwinger,
  Phys.\ Rev.\  {\bf 82}, 664 (1951).

\bibitem{Heinzl:2010vg} 
  T.~Heinzl, A.~Ilderton and M.~Marklund,
  Phys.\ Lett.\ B {\bf 692}, 250 (2010)

\bibitem{Schutzhold:2008pz} 
  R.~Schutzhold, H.~Gies and G.~Dunne,
  Phys.\ Rev.\ Lett.\  {\bf 101}, 130404 (2008)


\bibitem{PhysRevA.81.022122}
  G. R. Mocken, M. Ruf, C. M\"uller and C.H. Keitel,
  Phys.\ Rev.\ A {\bf 81}, 022122 (2010).
{
\bibitem{Titov:2012rd} 
  A.~I.~Titov, H.~Takabe, B.~Kampfer and A.~Hosaka,
  Phys.\ Rev.\ Lett.\  {\bf 108}, 240406 (2012)
}%
{
\bibitem{Nousch:2012xe} 
  T.~Nousch, D.~Seipt, B.~Kampfer and A.~I.~Titov,
  Phys.\ Lett.\ B {\bf 715}, 246 (2012).
}  
\bibitem{Smolyansky:1997fc} 
  S.~A.~Smolyansky, G.~Ropke, S.~M.~Schmidt, D.~Blaschke, V.~D.~Toneev and A.~V.~Prozorkevich,
  hep-ph/9712377.

\bibitem{Kluger:1998bm} 
  Y.~Kluger, E.~Mottola and J.~M.~Eisenberg,
  Phys.\ Rev.\ D {\bf 58}, 125015 (1998)
  
\bibitem{Schmidt:1998vi} 
  S.~M.~Schmidt, D.~Blaschke, G.~Ropke, S.~A.~Smolyansky, A.~V.~Prozorkevich and V.~D.~Toneev,
  Int.\ J.\ Mod.\ Phys.\ E {\bf 7}, 709 (1998)

\bibitem{Delone:1994}
N.~B.~Delone and V.~P.~Krainov, ``Multiphoton Processes in Atoms'', Springer (Berlin) (1994).


\bibitem{Popov:1973}
V.~S.~Popov, Zh.\ Eksp.\ Teor.\ Fiz.\ {\bf 63}, 1586 (1972) [Sov.
Phys. JETP 36, 840 (1973)].

\bibitem{Salieres:2001}
P. Salières \textit{ et al.}, 
Science {\bf 292}, 902 (2001).

\bibitem{Dunne:2005sx} 
  G.~V.~Dunne and C.~Schubert,
  Phys.\ Rev.\ D {\bf 72}, 105004 (2005)
  G.~V.~Dunne, Q.~-h.~Wang, H.~Gies and C.~Schubert,
  Phys.\ Rev.\ D {\bf 73}, 065028 (2006)
  
\bibitem{Muller:2009zzf} 
  C.~Muller, K.~Z.~Hatsagortsyan, M.~Ruf, S.~J.~Muller, H.~G.~Hetzheim, M.~C.~Kohler and C.~H.~Keitel,
  Laser Phys.\  {\bf 19}, 1743 (2009).
  
\bibitem{Ringwald:2001ib} 
  A.~Ringwald,
  Phys.\ Lett.\ B {\bf 510}, 107 (2001)
  C.~Kohlfurst, M.~Mitter, G.~von Winckel, F.~Hebenstreit and R.~Alkofer,
  Phys.\  Rev.\ D {\bf 88}, 045028 (2013)
  
\bibitem{Ruf:2009zz} 
  M.~Ruf, G.~R.~Mocken, C.~Muller, K.~Z.~Hatsagortsyan and C.~H.~Keitel,
  Phys.\ Rev.\ Lett.\  {\bf 102}, 080402 (2009)

\bibitem{Hebenstreit:2011wk} 
  F.~Hebenstreit, R.~Alkofer and H.~Gies,
  Phys.\ Rev.\ Lett.\  {\bf 107}, 180403 (2011)
  
\bibitem{Bloch:1999eu} 
  J.~C.~R.~Bloch, V.~A.~Mizerny, A.~V.~Prozorkevich, C.~D.~Roberts, S.~M.~Schmidt, S.~A.~Smolyansky and D.~V.~Vinnik,
  Phys.\ Rev.\ D {\bf 60}, 116011 (1999)
  
\bibitem{Alkofer:2001ik} 
  R.~Alkofer, M.~B.~Hecht, C.~D.~Roberts, S.~M.~Schmidt and D.~V.~Vinnik,
  Phys.\ Rev.\ Lett.\  {\bf 87}, 193902 (2001)
  
  \bibitem{Kobold:2002}
R.~Kopold, W.~Becker, M.~Kleber, G.~G.~Paulus,
J.\ Phys.\ B {\bf 35}, 217 (2002).
  
\bibitem{Brezin:1970xf} 
  E.~Brezin and C.~Itzykson,
  Phys.\ Rev.\ D {\bf 2}, 1191 (1970).
  
\bibitem{Hebenstreit:2013qxa} 
  F.~Hebenstreit, Jür.~Berges and D.~Gelfand,
  Phys.\ Rev.\ D {\bf 87}, 105006 (2013)
  
\bibitem{Hebenstreit:2009km} 
  F.~Hebenstreit, R.~Alkofer, G.~V.~Dunne and H.~Gies,
  Phys.\ Rev.\ Lett.\  {\bf 102}, 150404 (2009)

\end{thebibliography}

\end{document}